# U(1) flux tube profiles from Hamiltonian lattice gauge theory using a random walk ground-state projector

C. Best and A. Schäfer[a]*

[a]Institut für Theoretische Physik, J. W. Goethe-Universität, 60054 Frankfurt, Germany

We use a self-guided random walk to solve the ground-state problem of Hamiltonian U(1) pure gauge theory in 2+1 dimensions in the string sector. By making use of the electric-field representation, we argue that the spatial distribution of the electric field can be more easily measured than in ordinary Monte Carlo simulations.

## 1. INTRODUCTION

Recently, some focus has been given to the extraction of flux tube profiles from lattice gauge theory [1–3]. The usual approach involves calculating plaquette-plaquette correlators in Monte Carlo simulations from which the field distribution can be derived. Due to the high level of statistical fluctuations, this requires very large statistics and has only recently become feasible.

However, flux tubes and strings arise quite naturally from the strong-coupling expansion. We propose to use the Hamiltonian formulation of lattice gauge theory to calculate field-strength distributions. The Hamiltonian formulation has the advantage of not only having a simple strong-coupling limit but can be formulated in the electric-field basis, making the field strength a fundamental observable. Also, the Hamiltonian theory separates in different sectors according to string length, reducing the difficulty of extracting one particular sector.

## 2. METHOD

### 2.1. Electric-field representation

We attempt to solve the ground-state problem of the Kogut-Susskind Hamiltonian

$$\hat{H} = \frac{1}{2} \sum_{(ij)} \hat{l}_{ij}^2$$
$$+ \lambda \sum_p \left( 1 - \frac{1}{2} \operatorname{tr} \hat{U}_p - \frac{1}{2} \operatorname{tr} \hat{U}_p^+ \right) \quad .$$

*Work supported by Deutsche Forschungsgemeinschaft.

$\hat{l}_{ij}$ is the link operator (electric field) associated to the link $ij$, $\hat{U}_p$ the plaquette operator (magnetic field) of plaquette $p$, $\lambda \sim 1/g^2$ the coupling constant.

In the electric-field representation, the basis of the Hilbert space is chosen such that $\hat{l}_i$ is diagonal. The basis states of the representation are labeled by lattice configurations where each link carries a whole number $n_l$ of electric flux units. The plaquette operator (magnetic field) acts as a simple raising operator:

$$\langle n' | \hat{U}_p^+ | n \rangle = \delta_{n',n+p}$$

where $n + p$ represents the basis state $n$ with the flux around plaquette $p$ raised by one unit.

The Hamiltonian reads in this representation:

$$\hat{H} = \hat{H}_{\text{el}} + \hat{H}_{\text{mag}}$$
$$\langle n' | \hat{H}_{\text{el}} | n \rangle = \frac{1}{2} \delta_{n',n} \sum_{(ij)} n_{ij}^2$$
$$\langle n' | \hat{H}_{\text{mag}} | n \rangle =$$
$$= \lambda \sum_p \left( \delta_{n',n} - \frac{1}{2} \delta_{n',n+p} - \frac{1}{2} \delta_{n',n-p} \right)$$

Invariance under gauge transformations implies that the Hamiltonian preserves Gauss' law at each site, as the total flux at any site is not changed by the application of the plaquette operator $\hat{U}_p$. The Hilbert space separates into sectors of different external charges that act as sources of the electric field.

In the strong-coupling limit, $\lambda \to 0$ and $\hat{H}_{\text{mag}} \to 0$, and the Hamiltonian is diagonal. In each sector, the ground state is the shortest string linking the charges. Thus, string states are very natural in this model, and numerical simulation can center on how they spread out and acquire a finite thickness, and how they possibly vanish at a phase transition.

## 2.2. Random walks

We employ an ensemble ground-state projector using a guided random walk [4,5] to calculate the ground-state wave function. This method makes use of the similarity between the imaginary-time Schrödinger equation

$$\frac{\partial}{\partial t}|\psi_t\rangle = -\hat{H}|\psi_t\rangle$$

that governs the evolution of a state $|\psi_t\rangle$ towards the ground-state projector

$$\hat{P}_0 = \lim_{t\to\infty} e^{-t\hat{H}}$$

and an ensemble random walk with transition probabilities $P(n',n)$ and survival factor $V(n)$:

$$\psi_{t+\Delta t}(n') = V(n')\sum_n P(n',n)\psi_t(n) \quad .$$

In an ensemble random walk, members not only undergo transitions according to the $P(n',n)$ but are also deleted or replicated according to the survival factor $V(n)$.

For a lattice gauge Hamiltonian, ensemble members are lattice configurations $n$. The wave function $\psi_t(n)$ is approximated by the distribution of ensemble members in this space. It is driven towards the ground-state wave function by means of appropriate transition probabilities and survival factor $P(n',n)$ and $V(n)$ that are obtained by equating the discretized imaginary-time Schrödinger and the random-walk equation.

The probability interpretation of $P(n',n)$ and $V(n)$ restricts the applicability of the random-walk method to certain Hamiltonians. In particular, while $U(1)$ lattice gauge theory satisfies this restriction, nonabelian theories fail as they would require negative probabilities or, equivalently, do not ascertain that the ground-state wave function is positive.

## 2.3. Guidance

In a many-dimensional problem, the random walk easily strays into improbable regions of the basis space. Is it thus preferable to offer some guidance in the choice of "good" transitions. This is done by multiplying the Hamiltonian with a guidance wave function $\Phi(n)$:

$$\hat{\tilde{H}} = \Phi^{-1}\hat{H}\Phi$$
$$\langle n'|\hat{\tilde{H}}|n\rangle = \frac{1}{\Phi(n')}\langle n'|\hat{H}|n\rangle\Phi(n)$$

The guided Hamiltonian has the same spectrum as the original Hamiltonian; its wave functions are multiplied ("masked") by the guidance wave function. The guidance basically makes moves towards the guidance function more probable.

As the guidance changes the eigenfunctions, the extraction of observables requires either to use a guidance wave function that is sufficiently close to the actual wave function, or to use a split-time measurement that removes the guidance.

We chose a link-separable guidance function based on a modified Bessel function and parametrized by the center and the width of the distribution. As the ground state in the string sector is spatially inhomogeneous, different parameters can be assigned to each link. We introduce a form of self-guidance by constantly adapting the parametrization to the measured observables such that the ensemble described by the guidance wave function yields the same observables as the actual ground-state ensemble. In this sense, the method is variational: it evolves the parametrization of the guidance wave function towards an optimal parametrization.

The variational character of this method reduces artifacts from the guidance function as the result should be at least as good as one obtained from a purely variational method, which would be sufficient for measuring the field distribution. Is also hints at how the method can be extended to systems that are not easily represented as a random walk, such as nonabelian theories.

## 3. CALCULATIONS AND RESULTS

We are testing our method in $2+1$ dimensions on $U(1)$ lattice gauge theory. Typically, each ensemble has 50 members that are evolved over 200–300 steps. The ensemble size is held stable by a suitable energy offset. While the generation process is fast, high statistics (typical $10^6$ to $10^7$ configurations) is needed as the wave function acquires a large width outside the strong-coupling region. The self-guidance introduces another source of error, as one has to wait until the parametrization stabilizes.

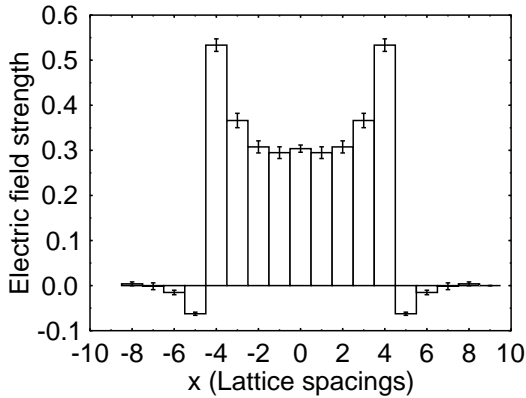

Figure 1. Electric field strength (in flux units) in a cut along the string axis (String of 9 on a $18 \times 11$ lattice at $\lambda = 1.5$). Errors are statistical.

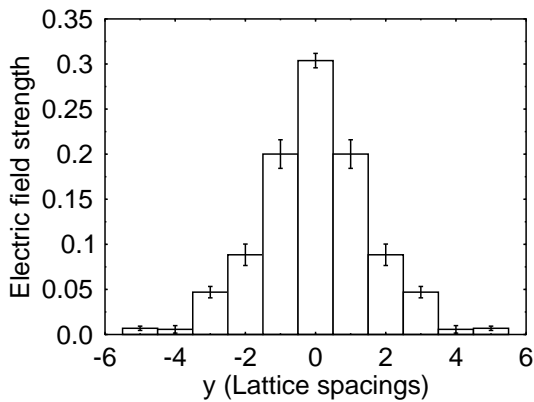

Figure 2. Electric field strength in a cut across the string axis.

Fig. 1 shows the distribution of electric field strengths in a cut along the string axis for a sample string. The deviation from strong coupling is manifested in the drop of the field at the center of the string accompanied by a spreading out shown in the transverse cut in Fig. 2.

While electric field strengths are easily accessible in this approach, it is somewhat harder to estimate the string tension. The vacuum energies dominate the ground-state energies by two orders of magnitudes such that calculating the energy difference is impracticable. However, we see some indication that the classical energy density $\langle \vec{E} \rangle^2$ may serve as a substitute observable.

## 4. SUMMARY

We presented the application of the guided random walk ensemble ground-state projection method to the calculation of field distributions in flux tubes. This methods targets directly the string sector of the theory and thus provides a more efficient way to calculate field strengths. In particular, they do not appear as plaquette-plaquette correlations but as primary representation of the wave function.

In the case of $2 + 1$–dimensional $U(1)$ gauge theory, one does not expect a phase transition to a nonconfined phase. For further investigations, we plan to consider the $3 + 1$–dimensional case in which a deconfinement phase transition will occur.

To apply the method to nonabelian theories, an appropriate scheme is worked at to use the random walk as a variational projector which will at least return a best-fit positive wave function.

Finally, inclusion of fermions is straightforward in the Hamiltonian formulation both as an approximation as well as exactly.

REFERENCES

1. W. Feilmair, H. Markum, Nucl. Phys. B370, 299 (1992).
2. Y. Peng, R. W. Haymaker, Phys. Rev. D47, 5104 (1993).
3. G. S. Bali, C. Schlichter, K. Schilling, Nucl. Phys. B (Proc. Suppl.) 34, 216 (1994).
4. S. A. Chin, J. W. Negele, S. E. Koonin, Ann. Phys. 157, 140 (1984).
5. S. E. Koonin, E. A. Umland, M. R. Zirnbauer, Phys. Rev. D33, 1795 (1986).